\documentclass[aps,prl,final,twocolumn,letterpaper]{revtex4}

\usepackage{graphicx}   
\usepackage{import}                         
\usepackage{epstopdf}
\usepackage{amsmath} 
\usepackage{float} 
\usepackage{bm}
\usepackage{amssymb}
\usepackage{quotes}
\usepackage{indentfirst}
\usepackage{color}
\usepackage{transparent}
\usepackage{dcolumn}
\usepackage{braket}
\usepackage{multirow}
\usepackage{cancel} 
\usepackage{mdframed}
\usepackage{color}
\usepackage{bm}
\usepackage{dsfont}
\usepackage{slashed}
\usepackage{soul, color} 
\soulregister\ref{7}  
\soulregister\cite{7} 
\renewcommand{\st}[1]{}

\usepackage{xr}

\makeatletter
\newcommand*{\addFileDependency}[1]{
  \typeout{(#1)}
  \@addtofilelist{#1}
  \IfFileExists{#1}{}{\typeout{No file #1.}}
}
\makeatother

\newcommand*{\myexternaldocument}[1]{%
    \externaldocument{#1}%
    \addFileDependency{#1.tex}%
    \addFileDependency{#1.aux}%
}

\myexternaldocument{si}

\usepackage{textcomp} 
\usepackage{xifthen}
\usepackage{xcolor}
\usepackage{etoolbox}
\newboolean{togglechanges} 

\setboolean{togglechanges}{false}

\newcommand{\comment}[1]{\ifbool{togglechanges}
    {#1}  
    {\textcolor{blue}{#1}}}

\usepackage{bibentry}

\usepackage{graphicx}
\usepackage{dcolumn}
\usepackage{bm}

\usepackage{textcomp} 

\begin{document}
\rmfamily


\title{Photonic probabilistic machine learning using quantum vacuum noise}
\author{Seou~Choi$^{1}$}
\email{seouc130@mit.edu}
\author{Yannick~Salamin$^{1,2}$}
\author{Charles~Roques-Carmes$^{1,3}$}
\email{chrc@stanford.edu}
\author{Rumen~Dangovski$^{1,4}$}
\author{Di~Luo$^{4,5,6}$}
\author{Zhuo~Chen$^{2,4}$}
\author{Michael~Horodynski$^{2}$}
\author{Jamison~Sloan$^{1}$}
\author{Shiekh~Zia~Uddin$^{1,2}$}
\author{Marin~Solja\v{c}i\'{c}$^{1,2}$}

\affiliation{$^{1}$ Research Laboratory of Electronics, MIT, Cambridge MA USA\looseness=-1}
\affiliation{$^{2}$ Department of Physics, MIT, Cambridge MA USA\looseness=-1}
\affiliation{$^{3}$ E. L. Ginzton Laboratories, Stanford University, 348 Via Pueblo, Stanford, CA USA\looseness=-1}
\affiliation{$^{4}$The NSF AI Institute for Artificial Intelligence and Fundamental Interactions, Cambridge, MA 02139, USA\looseness=-1}
\affiliation{$^{5}$ Center for Theoretical Physics, Massachusetts Institute of Technology, Cambridge, MA 02139, USA\looseness=-1}
\affiliation{$^{6}$Department of Physics, Harvard University, Cambridge, MA 02138, USA\looseness=-1}



\clearpage

\renewcommand{\sp}{\sigma_+}
\newcommand{\pbit}{$p$-bit}
\newcommand{\pbits}{$p$-bits}
\newcommand{\sm}{\sigma_-}
\newcommand{\mnist}{MNIST-handwritten digits}
\newcommand{\mnistnos}{MNIST-handwritten digit}

\setlength{\parskip}{0em}
\vspace*{-2em}


\vspace{0.8cm}

\begin{abstract}
    Probabilistic machine learning utilizes controllable sources of randomness to encode uncertainty and enable statistical modeling. Harnessing the pure randomness of quantum vacuum noise, which stems from fluctuating electromagnetic fields, has shown promise for high speed and energy-efficient stochastic photonic elements. Nevertheless, photonic computing hardware which can control these stochastic elements to program probabilistic machine learning algorithms has been limited. Here, we implement a photonic probabilistic computer consisting of a controllable stochastic photonic element -- a photonic probabilistic neuron (PPN). Our PPN is implemented in a bistable optical parametric oscillator (OPO) with vacuum-level injected bias fields. We then program a measurement-and-feedback loop for time-multiplexed PPNs with electronic processors (FPGA or GPU) to solve certain probabilistic machine learning tasks. We showcase probabilistic inference and image generation of \mnist, which are representative examples of discriminative and generative models. In both implementations, quantum vacuum noise is used as a random seed to encode classification uncertainty or probabilistic generation of samples. In addition, we propose a path towards an all-optical probabilistic computing platform, with an estimated sampling rate of $\sim 1~\text{Gbps}$ and energy consumption of $\sim \text{5~fJ}/\text{MAC}$. Our work paves the way for scalable, ultrafast, and energy-efficient probabilistic machine learning hardware.
\end{abstract}

\maketitle

Probabilistic machine learning can accelerate image generation \cite{nichol2021glide, goodfellow2014generative}, heuristic optimization \cite{roques2020heuristic, pham2012intelligent}, and probabilistic inference \cite{blundell2015weight, neal2012bayesian} by leveraging stochasticity to encode uncertainty and enable statistical modeling \cite{murphy2022probabilistic,ghahramani2015probabilistic}. These approaches are well suited for real-life applications which must account for uncertainty and variability, including autonomous driving \cite{feng2021review}, medical diagnosis \cite{richens2020improving}, and drug discovery \cite{jimenez2020drug}. However, digital complementary metal-oxide-semiconductor (CMOS) technology requires extensive resource overhead to simulate randomness and control probabilities, which leads to significantly increased power consumption and decreased operational speed \cite{qin2023high}. These challenges have sparked recent proposals for beyond-CMOS hardware such as low-barrier magnetic tunnel junctions~\cite{chowdhury2023full} and diffusive memristors~\cite{woo2022probabilistic} --- both of which leverage intrinsic noise as a source of randomness.

Concurrently, optical neural networks (ONNs) \cite{wetzstein2020inference,mcmahon2023physics} have shown remarkable progress in energy efficiency \cite{wang2022optical,ma2023quantum}, speed \cite{mourgias2022noise} and bandwidth \cite{totovic2022programmable} for solving deterministic tasks such as image classification \cite{bernstein2023single} and speech recognition \cite{shen2017deep}. Another important feature of ONNs is the inherent presence of noise in their operation. Therefore, photonic computing hardware typically implements computational tasks that are robust to optical noise~\cite{mcmahon2023physics}. ONNs have also been explored in regimes where deterministic tasks are performed with high accuracy, despite the presence of high levels of inherent noise~\cite{ma2023quantum}. Conversely, ONNs in which optoelectronic noise is intentionally added have also been proposed for optimization~\cite{prabhu2020accelerating} and generative networks~\cite{wu2022harnessing}. Meanwhile, optics has demonstrated high-data rate random number generation by harnessing quantum vacuum noise \cite{marandi2012all,kim2021massively}, and has shown natural abilities to generate probability distributions from quantum noise~\cite{roques2023biasing,wu2022photonic,ma2023stochastic}. However, the experimental demonstration of a photonic probabilistic machine learning system has remained elusive so far, mostly due to the lack of \textit{programmable} stochastic photonic elements.

Here, we experimentally demonstrate a probabilistic computing platform utilizing photonic probabilistic neurons (PPNs). Our PPN is implemented as a biased degenerate optical parametric oscillator (OPO), which leverages quantum vacuum noise to generate a probability distribution encoded by a bias field. We realized a hybrid opto-electronic probabilistic machine learning system which combines time-multiplexed PPNs and electronic processors with algorithm-specific measurement-and-feedback strategies. We demonstrate probabilistic inference of \mnist~with a stochastic binary neural network (SBNN), highlighting how quantum vacuum noise can encode classification uncertainty in discriminative models. Additionally, we showcase the generation of \mnist~with a pixel convolutional neural network (pixelCNN), demonstrating how statistical sampling in generative models can be facilitated by quantum vacuum noise. Furthermore, we provide a thorough discussion of the potential of an all-optical probabilistic machine learning system, offering a possible performance enhancement by a factor of 100 in both speed and energy over traditional CMOS implementations, thereby opening new avenues in high-speed, energy-efficient computing applications.

\begin{figure*}[t]
    \centering
    \includegraphics[scale = 0.8]{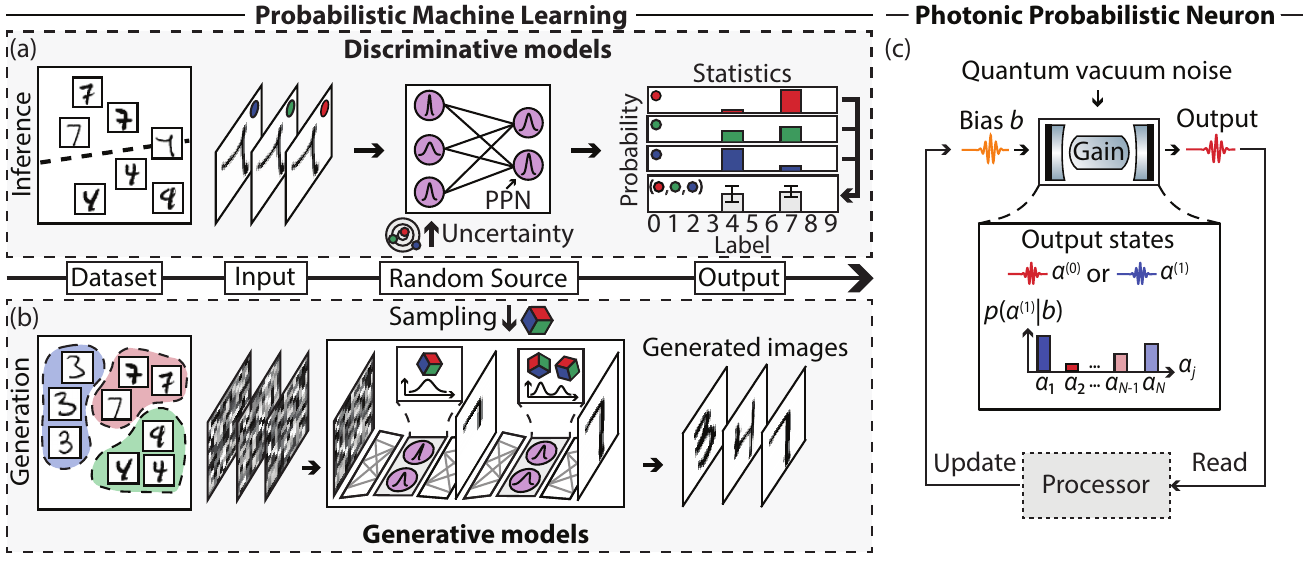}
    \vspace*{-4mm}
    \caption{\textbf{Probabilistic machine learning with stochastic photonic elements.} (\textbf{a} and \textbf{b}) Probabilistic machine learning enabled by physical random sources, solving (\textbf{a}) inference and (\textbf{b})~generation tasks. Neural networks learn a decision line for inference tasks and overall distribution for generation tasks. (\textbf{a})  
    Random sources encode uncertainty in neural network parameters, allowing statistical interpretation on inference results. (\textbf{b}) Stochastic image generation seeded by random sources samples new images from certain probability distributions stored in neural networks. Both computational tasks require controllable stochastic photonic elements which can learn probability distribution and perform statistically independent sampling, which we refer to as photonic probabilistic neurons (PPNs).  (\textbf{c}) Schematic of PPNs. One of the output states ($\alpha^{(0)}$ or $\alpha^{(1)}$) of a multistable optical system is randomly selected from a certain state probability distribution $p(\alpha^{(1)}|b)$ controlled by a bias field level \textit{b}. Subsequently, a processing unit reads the output state and updates the bias value $b$ for the next sampling. $N$ independent outcomes can be sampled from different  probabilities by time-multiplexing the bias signal.}
    \label{fig:probabilisitic computing}
\end{figure*}

\section{Results}
\subsection*{Probabilistic computing with time-multiplexed PPNs}

We first provide a brief overview of two probabilistic machine learning models and their optical implementation with PPNs (Fig.~\ref{fig:probabilisitic computing}).
 
Discriminative models learn decision lines that encode classification boundaries between different images (Fig.~\ref{fig:probabilisitic computing}(a),~left)~\cite{jebara2012machine}. Probabilistic neural networks (Fig.~\ref{fig:probabilisitic computing}(a),~middle) then impart statistical properties onto network parameters (e.g., weight uncertainty~\cite{blundell2015weight} or layer nodes~\cite{srivastava2014dropout}). Therefore, the network can provide a statistical ensemble of classification results, which are shown as different probabilities of the image classified to the certain labels (Fig.~\ref{fig:probabilisitic computing}(a), right). Probabilistic inference can quantify classification uncertainty, which becomes critical for ambiguous images located near the decision boundary \cite{silipo1998artificial,leibig2017leveraging}. 

On the other hand, generative models learn the underlying probability distribution of the training dataset (e.g., images) in order to create new ones (Fig.~\ref{fig:probabilisitic computing}(b), left)~\cite{jebara2012machine}. When generating new images, generative models use random sources to seed stochastic image sampling based on the probability distribution learned by the network (Fig.~\ref{fig:probabilisitic computing}(b), middle). As a result, images with different labels can be generated (Fig.~\ref{fig:probabilisitic computing}(b), right). 

In both of these computational tasks, probabilistic machine learning requires stochastic photonic elements whose probability distribution can be tuned, and that can perform statistically independent sampling. We refer to the optical implementation of this capability as PPNs (purple circles in Fig.~\ref{fig:probabilisitic computing}(a), and (b)). 

The proposed PPN is depicted in Fig.~\ref{fig:probabilisitic computing}(c). The building block consists of a synchronously pumped degenerate OPO~\cite{roques2023biasing}. An OPO consists of a nonlinear medium (e.g., second order nonlinear crystal, down converting photon frequency) and an optical cavity surrounding it. The phase of the initial optical state is random due to electromagnetic field fluctuations inside the cavity (quantum vacuum noise). When the power of the pump laser exceeds a certain threshold power, phase-sensitive gain of the OPO allows the initial state to fall into one of the bistable output states with either phase ($0~\mathrm{rad}$, or $\pi ~\mathrm{rad}$)~\cite{marandi2012all}. In other words, quantum vacuum noise acts as a perfect random source that manifests itself in the output state. If a vacuum-level external bias field $b$ is introduced in the OPO cavity, the probability distribution of the output steady states can be coherently controlled~\cite{roques2023biasing}. Specifically, our OPO-based PPN encodes a Bernoulli trial $B(p)$ with binary outcomes having probability $p$ and $1-p$. Independent random sampling and processing can be realized by time-multiplexing the bias signal, resulting in $N$ independent outcomes with encoded probabilities as depicted by different heights in Fig.~\ref{fig:probabilisitic computing}(c).

\begin{figure}[t]
    \centering
    \includegraphics[scale = 0.75]{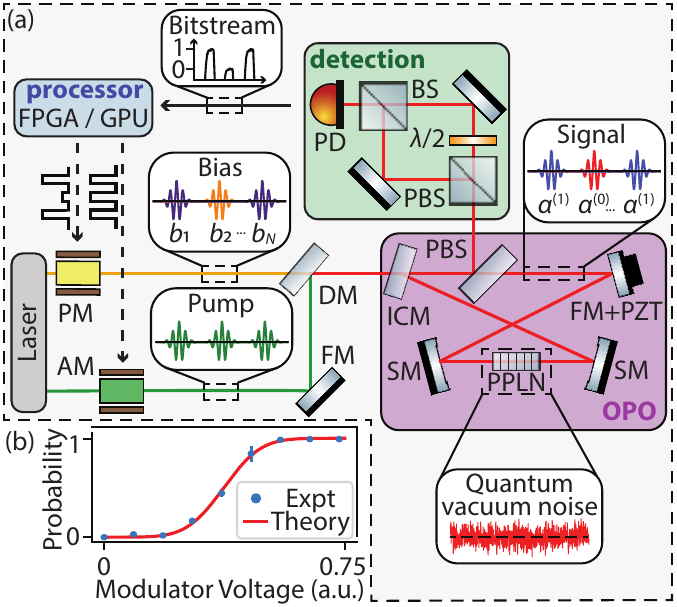}
    \vspace*{-4mm}
    \caption{\textbf{Experimental demonstration of a photonic probabilistic computer.}  (\textbf{a}) Experimental setup, consisting of an ultrafast laser pumping a nonlinear cavity, and homodyne detection to measure the phase of the OPO signal. Electronic processing units (FPGA/GPU) generate electrical signals to tune the probability. AM: amplitude modulator, PM: phase modulator, FM: flat mirror, DM: dichroic mirror, ICM: in-coupling mirror, (P)BS: (polarization) beam-splitter, SM: spherical mirror, PPLN: periodically poled lithium niobate nonlinear crystal. PZT: piezoelectric actuator, $\lambda/2$: half waveplate, PD: photodiode. (\textbf{b}) Modulator voltage - probability relationship. }
    \label{fig:setup}
\end{figure}

The experimental system realizing the PPN, and its implementation into a probabilistic computing system, is shown in Fig.~\ref{fig:setup}. The system consists of three modules: biased OPO (purple area), detection (green area), and processing unit (blue area). We time-multiplex OPO signals with an amplitude modulator along the pump path to sample multiple binary outputs from a single optical cavity at a rate of 10~kHz. This bit rate is chosen to ensure statistical independence of each PPN~\cite{marandi2012all, roques2023biasing}. We use a homodyne detector to measure the optical phase of the steady state and map it to the corresponding bit value (i.e., $0~\mathrm{rad}$ $\rightarrow$ 0 and $\pi ~\mathrm{rad}$  $\rightarrow$ 1). 

During each cycle, a bit is measured by the homodyne detector (value 0 or 1), conditioned on the bias value $b$. This bit, or a collection of bit values (``bitstream''), is then fed into an electronic processing unit to update the bias field value and sample the PPN in the next cycle. In our experiment the processing unit is taken as either a field-programmable gated array (FPGA) or a graphics processing unit (GPU). The FPGA is more adapted for real-time bitstream processing and control of the optical system, while the GPU can accelerate complex machine learning algorithms such as image generation at the cost of a slower system control.

Individual $p_{i}$ values are encoded in the phase of the bias field $b_{i}$ by applying a calibrated square-wave voltage to a phase modulator in the bias line path. The voltage-probability relation provided by the phase modulator is shown in Fig.~\ref{fig:setup}(b). This relation is used in the following computing experiments to control the bias voltage. A detailed description of the experimental setup is discussed in the Supplementary Information (SI), Section S1.

\subsection*{Photonic probabilistic computer for image classification}

\begin{figure*}
    \centering
    \includegraphics[scale = 0.8]{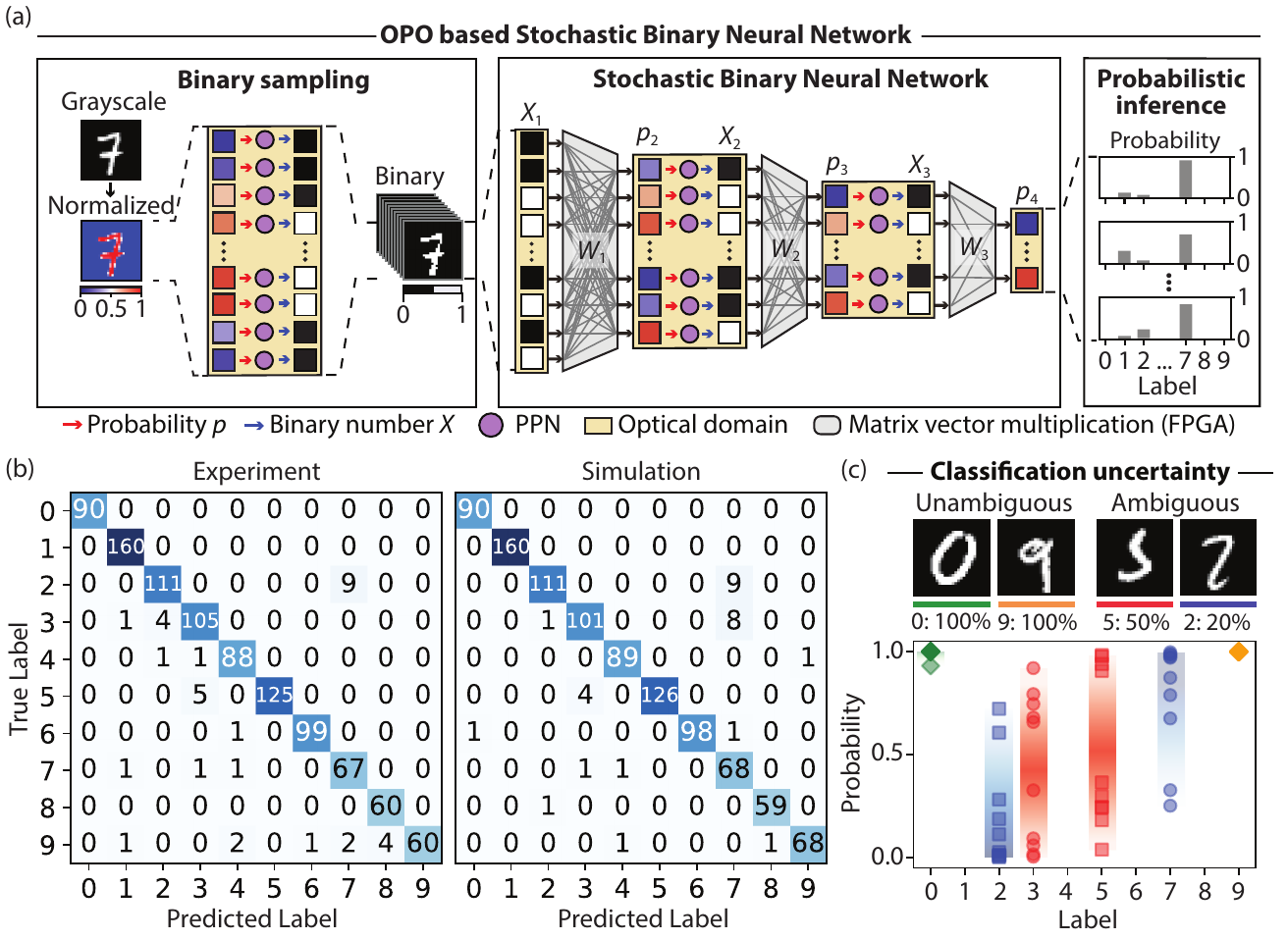}
    \vspace*{-4mm}    \caption{\textbf{Probabilistic inference uncertainty encoded by quantum vacuum noise} 
    (\textbf{a}) Hybrid photonic-electronic architecture for stochastic binary neural networks (SBNNs). Original MNIST grayscale handwritten-digit is binarized 10 times with PPNs and each binarized image propagates through SBNN (left panel). Binary nodes are sampled by PPNs and their corresponding \textit{p} values are evaluated by FPGA (middle panel). Because the nodes are stochastic, inference results vary (right panel). (\textbf{b}) Confusion matrices of image classification results. A total of 1,000 binary images ($100$ grayscale testing images  $\times$ 10 times of binarization $=1,000$ input images) are tested. (\textbf{c}) Diagnosing inference results with the aid of quantum vacuum noise. Breadth in probability and low classification accuracy reflect the ambiguity of the input image.}
    \label{fig:SBNN}
\end{figure*}

We now perform probabilistic image classification of \mnist~using a pre-trained SBNN model on our optical probabilistic computing platform (Fig.~\ref{fig:SBNN}(a)). SBNN encodes inference uncertainty by substituting deterministic layer nodes (as found in conventional fully-connected neural networks) with stochastic binary nodes \cite{li2023binary}. In a conventional, fully-connected neural network, the $j^\text{th}$ node value in the ${(n+1)}^\text{th}$ layer $X_{j,n+1}$ can be calculated in two steps: (1) matrix-vector multiplication (MVM) between weight matrix $W$ and the $n^\text{th}$ layer $X_{n}$ ($z_{j,n}\equiv\sum_{i}W_{j,i}X_{i,n}$); followed by (2) a nonlinear activation function 
$\sigma(\cdot)$ : $X_{j,n+1}=\sigma\left(z_{j,n}\right)$.

Within our SBNN model, each layer node is represented by a PPN, and a single layer (yellow areas in Fig.~\ref{fig:SBNN}(a)) is described as a bitstream of time-multiplexed PPNs. Because of the nonlinear nature of the bias-probability relationship (Fig.~\ref{fig:setup}(b)), sampling a binary output $X_{j,n}$ with our PPN from the given bias $b_{j,n}$ (or equivalently bias modulator voltage $V_{j,n}$ in our experiment), naturally corresponds to passing a nonlinear activation function: $X_{j,n}=B(p_{j,n})=B\left[\sigma\left(V_{j,n}\right)\right]$. Modulator voltage $V_{j,n}$ is calculated via MVM between the weight $W_{n-1}$ and the $\left( {n-1}\right)^\text{th}$ layer $X_{n-1}$ (gray areas in Fig.~\ref{fig:SBNN}(a), which is performed by the FPGA in our experiment). In other words, each PPN node binarizes the input, which consists of a weighted sum of previous layer nodes, with probability $p_{j,n}$. Because of the stochastic nature of the nodes, their probabilities change for every inference, leading to a probabilistic interpretation of classification results for an identical input image (right panel of Fig.~\ref{fig:SBNN}(a)). 

To perform image classification of \mnist~with our optical SBNN, we first binarize original \mnist~(left panel of Fig.~\ref{fig:SBNN}(a)). Original \mnist~(grayscale, pixel values ranging from 0 to 255) are normalized between 0 and 1. The resulting pixel values represent the probability value for each PPN. The grayscale images are then binarized by sampling the PPNs. The binary images are propagated through the network ($784 \rightarrow 128 \rightarrow 64 \rightarrow 10$), with real-time communication between PPNs and the FPGA. The output layer $O_{0,1,...,9}$ is used to interpret the classification result, higher $O_{j}$ corresponding to the higher probability of image representing digit ``$j$.'' The network is pre-trained \textit{in silico} and the weights are implemented on the FPGA. A detailed description of the training process and how the FPGA communicates with the optical setup is in the SI, Section S2. 

To test the performance of our optical SBNN, a batch of grayscale \mnist~(100 images) from the test set is selected. By binarizing each grayscale \mnistnos~10 times to encode statistical uncertainty, we prepared 1,000 binarized \mnist~ in total to be classified by our optical SBNN. Confusion matrices in Fig.~\ref{fig:SBNN}(b) show that the overall experimental classification accuracy (96.5\%) is in close agreement with the accuracy obtained from the numerical simulations (97.0\%).

\begin{figure*}
    \centering
    \includegraphics[scale = 0.8]{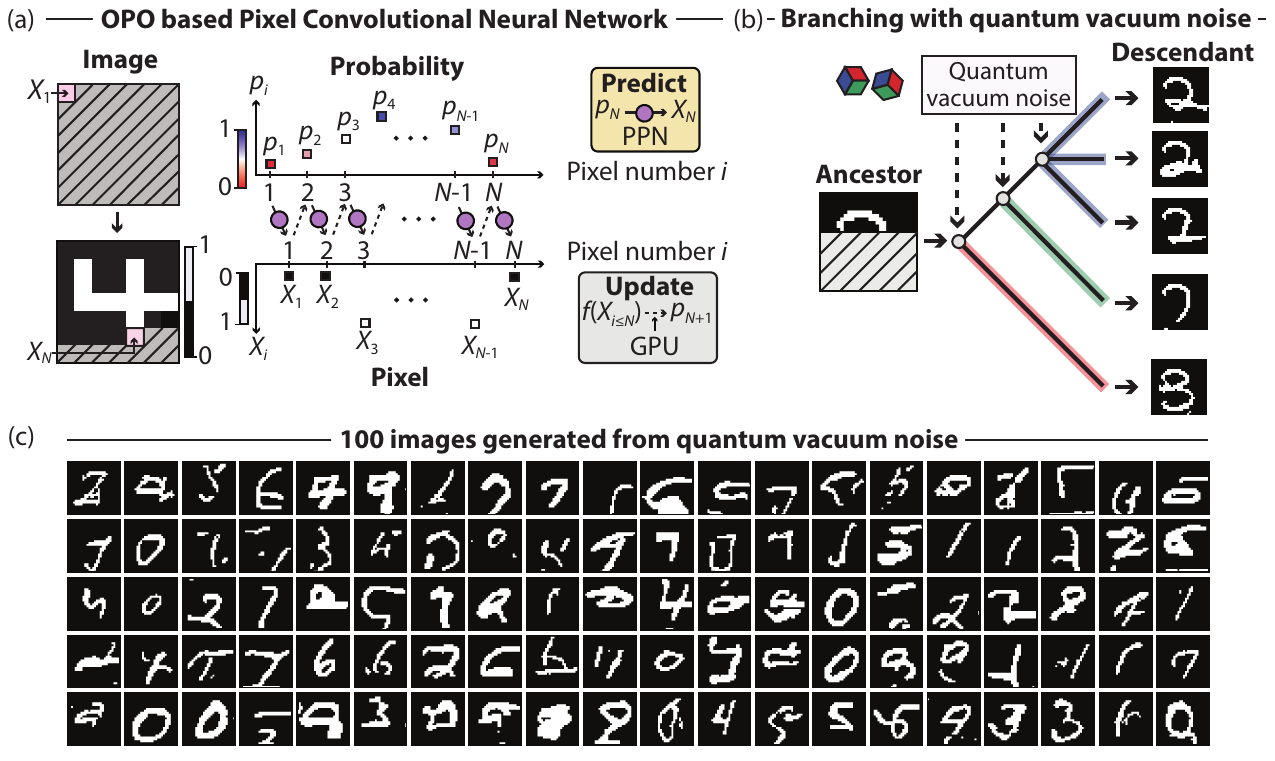}
    \vspace*{-4mm}    
    \caption{\textbf{Images sampled from quantum vacuum noise} (\textbf{a}) PixelCNN generating binary MNIST handwritten-digits. PPNs sample the pixel value $X_{N}$ from the given $p_{N}$ value and GPU calculates the $p_{N+1}$ value for the next pixel from the previous pixels $X_{i\leq N}$. (\textbf{b}) Branching off to different MNIST handwritten-digits, guided by quantum vacuum noise. Stochastic sampling allows generation of images with different digits and features. (\textbf{c}) 100 generated images from pixelCNN, starting from a complete empty image.} 
    \label{fig:pixelCNN}
\end{figure*}

Fig.~\ref{fig:SBNN}(c) shows how our probabilistic neural network can diagnose the reliability of inference results by harnessing quantum vacuum noise. Unlike deterministic neural networks, the variability of layer nodes in SBNNs results in different probability for each inference. One of the factors that can potentially degrade the classification performance is the ambiguity of the image (i.e., how close the image is to the decision boundary, as shown in Fig.~\ref{fig:probabilisitic computing}(a)). Three different scenarios are described in Fig.~\ref{fig:SBNN}(c). Unambiguous images such as ``0'' and ``9'' (achieving 100\% of classification accuracy) show relatively consistent classification results with probabilities of correct classification close to 1. In this scenario, probabilistic neural networks show similar behavior to deterministic neural networks, which always give the same classification result with a fixed probability. When the input image becomes ambiguous (image ``5'' underlined in red, achieving 50\% of classification accuracy), our SBNN model indicates that the image can be either ``3'' or ``5''. Accordingly, the distribution of probabilities on each label broadens with its average value close to 50\%. The worst case scenario is depicted by image ``2'' (underlined in blue), showing low overall accuracy (20\%) and strong inconsistency in classification results. Such scenario clearly showcases how probabilistic sampling can provide additional information to the end-user.

Offering both overall accuracy and statistics of classification results, probabilistic neural networks can diagnose inference results by providing a confidence level of the decision. The total classification result for each input image can be found in SI, Section S2.

\subsection*{Generating images from quantum vacuum noise with photonic generative models}

We now turn to the demonstration of generative models with our photonic probabilistic computing platform (Fig.~\ref{fig:pixelCNN}), demonstrating the use of quantum optical randomness as a source for generative machine learning models. We use a type of autoregressive model (pixelCNN), which models a conditional probability of a current pixel value from previous pixels~\cite{van2016pixel}. 

Our implementation protocol for pixelCNN with PPNs is described in Fig.~\ref{fig:pixelCNN}(a). A binary image with the first $N-1$ pixels $X_{i \leq N-1}$ specified is given as an input to the network. In principle, $N$ can be any natural number, $N=1$ corresponding to the case when pixelCNN creates an image only using quantum vacuum noise as a random seed. When the input image is given, a pre-trained pixelCNN model in the GPU evaluates $p_{N}$ to be encoded on the PPN from previous pixels $X_{i \leq N-1}$, generating a binary number for the $N^\text{th}$ pixel ($X_{N})$. The probability $p_{N+1}$ is now computed based on previous pixel values $X_{i \leq N}$. This process is repeated until the full image is generated ($28\times28 = 784$~pixels). Our hybrid opto-electronic computing system can generate new images using quantum vacuum noise as a random seed. Details of network structure and training method can be found in the SI, Section S3.

Different \mnist, all generated from the same incomplete input image, highlight how quantum vacuum noise enables stochastic image sampling (Fig.~\ref{fig:pixelCNN}(b)). Although they all start from the same  ``ancestor'' image, the multiple stochastic samples of pixel values from the PPNs branch off into different \mnist~with different labels (``descendant'' images). It is also possible to generate different images with the same label (which is likely to be labeled as ``2''). 

We produced 100 examples of handwritten digit images from quantum vacuum noise using our photonic probabilistic computing platform (Fig.~\ref{fig:pixelCNN}(c)). This was done by initiating an empty image as an input to our optical pixelCNN. We also test the negative-log-likelihood (NLL) of the generated images NLL $\equiv-\sum_i \{X_{i}\ln (p_{i})+(1-X_{i})\ln(1-p_{i}) \}$, where the sum runs over $i=1, \ldots, 784$ pixel indices. A lower value of NLL indicates statistical similarity to the distribution of training images, yielding $71.1 \pm 18.8$ for our experimental results and $64.9 \pm 15.4$ for numerical simulations. This shows that our system has learned an accurate representation of the image distribution. 

\section{Discussion}
In our demonstration of photonic probabilistic machine learning, the speed and energy efficiency were limited by the PPN sampling rate and data transfer bandwidth between electronic processors and PPNs. In the following, we propose an all-optical probabilistic computing platform which can overcome these challenges, and evaluate the potential benefit in terms of speed and energy efficiency compared to electronic state of the art. 

To increase sampling rate and reduce energy consumption, we propose an all-optical implementation. For instance, PPNs can be implemented with injection-seeded vertical-cavity surface-emitting lasers, reaching $>1~\text{Gbps}$~\cite{zhao2022fast} and providing energy-efficient operation \cite{chen2023deep}. Fast control of the probability and state detection can be achieved with high-bandwidth modulators and detectors \cite{valdez2023100,he2019high,wang2018integrated,lischke2015high,lischke2021ultra}, suggesting that PPNs achieving $1~\text{Gbps}$ sampling rate are within reach (detailed explanations can be found in SI, Section S4). 

Furthermore, our programmable stochastic element naturally implements an all-optical nonlinearity through the bias-probability relationship, which has been a historical challenge in the implementation of energy-efficient all-optical ONNs~\cite{wetzstein2020inference}. Typically, ONNs rely on optoelectronic measurement-feedback schemes to update the network layers~\cite{zhang2021optical, ashtiani2022chip}. Conversely, in the proposed scheme, an optical signal (\textit{vacuum-level bias}) controls the nonlinearity of the layer. Because the bias signal can be derived directly from the accumulated PPN outputs, bypassing active components, the scheme can reduce energy consumption per multiply-accumulate (MAC) operation to as low as $\sim \text{5~fJ}/\text{MAC}$.
State-of-the-art stochastic electronic devices, such as low-barrier magnetic tunnel junctions and diffusive memristors integrated with conventional CMOS technologies are expected to achieve $\sim 0.1~\text{Gbps}$~\cite{jiang2017novel,chen2022magnetic} and consume $\sim \text{900~fJ}/\text{MAC}$~\cite{li2023binary}. Comparatively, our proposed photonic platform can be $\sim \times 10$ faster and $\sim \times 100$ more energy efficient. A detailed discussion of this all-optical probabilistic computing platform is found in the SI, Section S4. 

We now compare the speed and energy performance of our photonic platform to a state-of-the-art FPGA~\cite{qin2020binary, umuroglu2017finn}, in an image classification task considering a binary neural network. The deterministic FPGA implementation demonstrated classification of $\sim 1.6$ million images per second with $\sim{23~\text{W}}$ power consumption. Adopting the network structure of our SBNN model in Fig.~\ref{fig:SBNN}, we can calculate the computation time and the number of MAC operations required for each inference. Our estimation gives $\sim~4~\text{ns}$ and $\sim~10^{5}$ MAC operations per classification, which result in $\sim$~250 million image classifications per second with a power consumption of $\sim\text{0.1~W}$. Therefore, the suggested all-optical probabilistic computing hardware could perform $\times$~100 faster while consuming $\times~100$ less power. Detailed discussion can be found in the SI, Section S4.

One of the possible extensions of our work is to train the network physically~\cite{wright2022deep,momeni2023backpropagation}. This becomes critical when an accurate digital modeling of the physical system becomes challenging due to its complexity. Without an additional cost of simulating randomness in digital models, several training methods which resort to stochasticity, including stochastic gradient descent~\cite{bottou2010large}, dropout~\cite{srivastava2014dropout}, and noise injection~\cite{zur2009noise} could potentially be realized with PPNs. Harnessing quantum vacuum noise in optical elements for both training and testing, our PPNs will pave the way of implementing all-optical probabilistic physical neural networks, which can benefit state-of-the-art machine learning applications including large language models~\cite{naveed2023comprehensive} and diffusion models~\cite{ho2020denoising}.  

\section{Online methods}

All the data and codes that are used within this paper are available from the corresponding authors upon request. Correspondence and requests should be addressed to S.~C. (seouc130@mit.edu) and C.~R.-C. (chrc@stanford.edu).

\section{Authors contributions}
S.~C., Y.~S., C.~R.-C., J.~S., and M.~S. conceived the original idea. R.~D. and D.~L. contributed to the development of the machine learning algorithms. S.~C. and Y.~S. built the experimental setup with contributions from C.~R.-C., M.~H., and S.~Z.~U.;~S.~C. acquired and analyzed the data. S.~C. developed the code for the electronic processing unit and training the neural networks with contributions from R.~D., D.~L., and Z.~C.; M.~S. supervised the project. The manuscript was written by S.~C., Y.~S., and C.~R.-C., with inputs from all authors.

\section{Competing interests}
The authors declare no potential competing financial interests.

\section{Acknowledgements}
The authors thank Wei Wang (Peng Cheng Laboratory) for providing resources used during the experimental demonstration of SBNNs. The authors also acknowledge Joe Steinmeyer for helpful discussions on FPGA. S.~C. acknowledges support from Korea Foundation for Advanced Studies Overseas PhD Scholarship. Y.~S. acknowledges support from the Swiss National Science Foundation (SNSF) through the Early Postdoc Mobility Fellowship No.~P2EZP2-188091. C.~R.-C. is supported by a Stanford Science Fellowship. D.~L., Z.~C., and M.~S. acknowledge support from the National Science Foundation under Cooperative Agreement PHY-2019786 (The NSF AI Institute for Artificial Intelligence and Fundamental Interactions, \url{http://iaifi.org/}). M.~H. acknowledges funding by the Austrian Science Fund (FWF) through grant J4729. J.~S. acknowledges earlier support from a Mathworks Fellowship. This material is based upon work supported by the U.S. Department of Energy, Office of Science, National Quantum Information Science Research Centers, Co-design Center for Quantum Advantage (C2QA) under contract number DE-SC0012704. This material is also based upon work sponsored in part by the U.S. Army DEVCOM ARL Army Research Office through
the MIT Institute for Soldier Nanotechnologies under Cooperative Agreement number W911NF-23-2-0121.

\bibliography{bibliography.bib}

\end{document}